\documentclass[aps,amsmath,amssymb]{revtex4}
\usepackage{graphicx}

\newcommand{\ba}{\begin{eqnarray}} \newcommand{\ea}{\end{eqnarray}}
\newcommand{\be}{\begin{equation}} \newcommand{\ee}{\end{equation}}
\newcommand{\bw}{\begin{widetext}} \newcommand{\ew}{\end{widetext}}

\renewcommand{\d}{\textrm{d}} 
\newcommand{\diag}{\textrm{diag}} \newcommand{\e}{\textrm{e}}
 \renewcommand{\i}{\textrm{i}}
 
 \renewcommand{\r}{\rho}
\newcommand{\s}{\sigma}
\renewcommand{\e}{\epsilon}\newcommand{\g}{\gamma}
 \renewcommand{\l}{\lambda}

\renewcommand{\bf}{\bfseries}
\newcommand{\fr}[2]{{\textstyle{\frac{#1}{#2}}}}
\renewcommand{\fr}{\tfrac}

\begin{document}

\title{Eccentric inflation and WMAP}\thanks{The calculations
presented in this talk (at Coral Gables 2003, Fort Lauderdale, FL)
appeared in short form in Ref.~[1] and a full exposition is to appear
in Ref.~[2].}

\author{Roman V. Buniy}
\email{roman.buniy@vanderbilt.edu}
\affiliation{Department of Physics and Astronomy, Vanderbilt
University, Nashville, TN 37235, USA}

\begin{abstract}
For uniform arrangements of magnetic fields, strings, or domain walls
(together with the cosmological constant and non-relativistic matter),
exact solutions to the Einstein equations are shown to lead to a
universe with ellipsoidal expansion. We argue the results can be used
to explain some features in the WMAP data. The magnetic field case is
the easiest to motivate and has the highest possibility of yielding
reliable constraints on observational cosmology.
\end{abstract}

\maketitle

\section{Introduction}\label{S:introduction}

The first year WMAP results
\cite{Bennett:2003bz,Spergel:2003cb,Hinshaw:2003ex} contain
interesting large-scale features which warrant further attention
\cite{Tegmark:2003ve,deOliveira-Costa:2003pu}. One glaring feature is
the suppression of power at large angular scales ($\theta \gtrsim
60^{\circ}$), which is reflected most distinctly in the reduction of
the quadrupole $C_2$ and octopole $C_3$. This effect was also seen in
the COBE results~\cite{Smoot:1992td,Kogut:1996us}.  While one can
argue that such an effect could just be statistical, it does not seem
unreasonable to try to model such behavior by altering the
cosmological model from the standard big bang plus inflation
scenario. More intriguingly, the quadrupole $C_2$ and octupole $C_3$
are found to be aligned.  In particular, the $l=2$ and $3$ powers are
found to be concentrated in a plane P inclined about $30^{\circ}$ to
the Galactic plane.  In a coordinate system in which the equator is in
the plane P, the $l= 2$ and $3$ powers are primarily in the $m=\pm l$
modes. The axis of this system defines a unique ray and supports an
idea of power on the axis is suppressed relative to the power in the
orthogonal plane. These effects seem to suggest one (longitudinal)
direction may have expanded differently from the other two
(transverse) directions in P.

We approach the issue of global anisotropy of the Universe by a simple
modification of the conventional Friedman-Robertson-Walker (FRW)
model. To achieve this, one has to consider an energy-momentum tensor
which either is spatially non-spherical or spontaneously becomes
non-spherical. Such a situation could occur when magnetic
fields~\cite{Kronberg:1993vk} or cosmic
defects~\cite{Hindmarsh:1994re} are present. Moreover, it is known
that large scale magnetic fields exist in the universe, perhaps up to
cosmological scales ~\cite{Kronberg:1993vk,Wick:2000yc}.

As a modest step toward understanding the form, significance and
implications of an asymmetric universe, we will modify the standard
spherically symmetric FRW cosmology to a form with only planar
symmetry \cite{Taub:1950ez}. Our choice of energy-momentum will result
in non-spherical expansion from an initially spherical symmetric
configuration: an initial co-moving sphere will evolve into an
ellipsoid that can be either prolate or oblate depending on the choice
of matter content.

For the sake of clarity, we first give general equations for
cosmologies with planar symmetry, (The universe looks the same from
all points but they all have a preferred direction.) and then
investigate one case in detail---a universe filled with dust, uniform
magnetic fields and cosmological constant.  This is perhaps the most
easily motivated, exactly solvable case to consider and it will give
us a context in which to couch the discussion of other examples with
planar symmetry and cases where planar symmetry is broken.

To set the stage, consider an early epoch in the universe at the onset
of cosmic inflation, where strong magnetic fields have been produced
in a phase transition
\cite{Savvidy:1977as,Matinian:1976mp,Vachaspati:nm,Enqvist:1994rm,Berera:1998hv}.
Assuming the magnitude of the magnetic field and vacuum energy
($\Lambda$) densities initially are about the same, we will find that
eventually $\Lambda$ dominates. It was estimated~\cite{Vachaspati:nm}
that the initial magnetic field energy produced in the electroweak
phase transition was within an order of magnitude of the critical
density. Other phase transitions may have even higher initial field
values \cite{Enqvist:1994rm,Berera:1998hv}, or high densities of
cosmic defects. Hence, it is not unphysical to consider a universe
with B-fields and $\Lambda$ of comparable magnitudes. If the B-fields
are aligned in domains, then some degree of inflation is sufficient to
push all but one domain outside the horizon.

\section{Planar geometry}\label{S:geometry}

To make the simplest directionally anisotropic universe, we modify FRW
spherical symmetry of space-time into planar symmetry; the most
general form for the metric then is~\cite{Taub:1950ez}
$(g_{\mu\nu})=\diag{(1,-e^{2a},-e^{2a},-e^{2b})}$, where $a$ and $b$
are functions of $t$ and $z$; the $xy$-plane is the plane of
symmetry. We also impose translational symmetry along the $z$-axis;
the functions $a$ and $b$ now depend only on $t$.

(Examples of planar-symmetric spaces include space uniformly filled
with either uniform magnetic fields, static alligned strings, or
static stacked walls; i.e., the defects are at rest with respect to
the cosmic backgound frame. This situation with defects is artificial
or at best contrived, but could perhaps arise in brane world physics
where static walls could be static or walls beyond the horizon could
be connected by static strings. We will not pursue these details
here. Of course, to these any spherically-symmetric contributions can
be added: vacuum energy, matter, radiation.)

To support the symmetry of space-time, the energy-momentum tensor for
the matter has to have the same symmetry, $({T^\mu}_\nu)=(8\pi G)^{-1}
\diag(\xi,\eta,\eta,\zeta)$. Here the energy density $\xi$,
transversal $\eta$ and longitudinal $\zeta$ tension densitites are
functions of time only.  The corresponding Einstein equations are
\ba&&\dot{a}^2+2\dot{a}\dot{b}=\xi,\label{A1}\\
&&\ddot{a}+\ddot{b}+\dot{a}^2+\dot{a}\dot{b}+\dot{b}^2=\eta,\label{A2}\\
&&2\ddot{a}+3\dot{a}^2=\zeta.\label{A3}\ea We also need the equation
expressing covariant conservation of the energy-momentum [a direct
consequence of Eqs.~(\ref{A1})--(\ref{A3})]: \ba
\dot{\xi}+2\dot{a}(\xi-\eta)+\dot{b}(\xi-\zeta)=0.
\label{A4}\ea 
For an adiabatic process, the entropy in a comoving volume is
conserved. For the matter with the equation of state $p=w\rho$, this
leads to\,\footnote{The subsript ``$\i$'' refers to the moment of
transition from isotropic to anisotropic dynamics.}
$\rho=\rho_\i\,e^{-(1+w)(2a+b)}$ and, as a result, the anisotropic
part of energy-momentum is conserved as well:
\begin{equation}
\dot{\tilde{\xi}} +2\dot{a}(\tilde{\xi}-\tilde{\eta})
+\dot{b}(\tilde{\xi}-\tilde{\zeta})=0.\label{A6}
\end{equation}
For the cases of magnetic field, strings and walls, the set of
quantities $(\tilde{\xi},\tilde{\eta},\tilde{\zeta})$ take,
respectively, the values $(\e,-\e,\e)$, $(\e,0,\e)$ and $(\e,\e,0)$.

Various relations and inequalities based on energy conditions, many of
which involve the eccentricity of the expansion can be derived for
general $\xi$, $\eta$ and $\zeta$; for details, see
Ref.~{\cite{long-version}}.

\section{Magnetic field}\label{S:B-field}

Extending the analysis of Ref.~\cite{Berera:2003tf}, we now present
the exact solution for the case of cosmological constant, dust ($w=0$)
and uniform B-field~\cite{long-version}.

Conservation of the anisotropic part of the energy-momentum,
Eq.~(\ref{A6}), gives $a=\fr{1}{4}\ln{(\e_\i/\e)}$. Using this result
in Eq.~(\ref{A3}), we find that $\e(t)$ is given implicitly by 
\begin{equation}
t-t_\i=\fr{1}{4} \int^{\e_\i}_{\e}\d\e\left[
\fr{1}{3}\l\e^2+\fr{1}{3}\e_\i^{-\frac{3}{4}} (4\e_\i+\rho_\i)
\e^{\frac{11}{4}}-\e^3\right]^{-\frac{1}{2}}.\label{B:integral}
\end{equation}
Eq.~(\ref{A1}) and $\rho=\rho_\i\,e^{-(2a+b)}$ give
\begin{eqnarray}
b=\fr{1}{2}\ln\frac{\l\e^{-\frac{1}{2}}
+(4\e_\i+\rho_\i)\e_\i^{-\frac{3}{4}}\e^{\frac{1}{4}}-3\e^{\frac{1}{2}}}
{(\l+\rho_\i)\e_\i^{-\frac{1}{2}}+\e_\i^{\frac{1}{2}}}
+\ln\left[1+F_\textrm{m}(\e)\right],\label{B:b}
\end{eqnarray}
where
\begin{eqnarray}
F_\textrm{m}(\e)=\frac{3\rho_\i}{8\e_\i}
\left(1+\frac{\l+\rho_\i}{\e_\i}\right)^{\frac{1}{2}}
\int_{\e/\e_\i}^{1}\d x\,x^{-\frac{1}{4}}
\left[\frac{\l}{\e_\i}+\left(4+\frac{\rho_\i}{\e_\i}\right)
x^{\frac{3}{4}}-3x\right]^{-\frac{3}{2}}.
\label{B:F}
\end{eqnarray}
 
A careful inspection of the above solution shows that space is oblate,
$e^{a-b}\ge 1$, and its oblateness monotonically increases from its
initial value (unity) to its asymptotic value. More magnetic field
increases the anisotropy; more matter reduces anisotropy, but neither
can change an oblate ellipsoid into a prolate one.

The asymptotics for large $t$ are as follows:
\begin{eqnarray}
&&\e\sim \e_\i\,
e^{-4(\l/3)^{\frac{1}{2}}(t-t_\i+\tau)},\label{B:e-asympt}\\ &&a\sim
(\l/3)^{\frac{1}{2}}(t-t_\i+\tau),\label{B:a-asympt}\\ &&b\sim
(\l/3)^{\frac{1}{2}}(t-t_\i+\tau)-\fr{1}{2}
\ln{\left[1+(\e_\i+\rho_\i)/\l\right]}+\ln\left[1+F_\textrm{m}(0)\right],
\label{B:b-asympt}\\
&&\rho\sim\rho_\i[1+(\e_\i+\rho_\i)/\l]^{\frac{1}{2}}[1+F_\textrm{m}(0)]^{-1}
e^{-(3\l)^{\frac{1}{2}}(t-t_\i+\tau)},\label{B:rho-asympt}
\end{eqnarray}
where
\begin{equation} 
\tau=\fr{1}{4}\int_{0}^{\e_\i}
\d\e\left\{\left(\fr{1}{3}\l\e^2\right)^{-\frac{1}{2}}
-\left[\fr{1}{3}\l\e^2+\fr{1}{3}\e_\i^{-\frac{3}{4}} (4\e_\i+\rho_\i)
\e^{\frac{11}{4}}-\e^3\right]^{-\frac{1}{2}}\right\}.\label{B:tau}
\end{equation}

\section{Strings}\label{S:strings}

From the conservation equation there follows
$a=\fr{1}{2}\ln{(\e_\i/\e)}$, and then Eq.~(\ref{A3}) gives $\e(t)$
implicitly via
\begin{equation} 
t-t_\i=\fr{1}{2} \int^{\e_\i}_{\e}\d\e\left[
\fr{1}{3}\l\e^2+\e^3+\fr{1}{3}\e_\i^{-\frac{3}{2}}(\rho_\i-2\e_\i)
\e^{\frac{7}{2}}\right]^{-\frac{1}{2}}.\label{S:integral}
\end{equation} 
Similarly to the case of the magnetic field, we solve Eq.~(\ref{A1})
and find
\begin{equation} 
b=\fr{1}{2}\ln\frac{3+\l\e^{-1}+(\rho_\i-2\e_\i)\e_\i^{-\frac{3}{2}}
\e^{\frac{1}{2}}} {1+(\l+\rho_\i)\e_\i^{-1}}+\ln\left[1+F_\textrm{s}(\e)\right],
\label{S:b}
\end{equation} 
where 
\begin{equation} 
F_\textrm{s}(\e)=\frac{3\rho_\i}{4\e_\i}
\left(1+\frac{\l+\rho_\i}{\e_\i}\right)^{\frac{1}{2}}
\int_{\e/\e_\i}^{1}\d x\,x^{\frac{1}{2}}
\left[\frac{\l}{\e_\i}+3x+\left(\frac{\rho_\i}{\e_\i}-2\right)
x^{\frac{3}{2}}\right]^{-\frac{3}{2}}.
\label{S:F}
\end{equation} 

A careful inspection of the above solution shows that space is oblate,
$e^{a-b}\ge 1$, and its oblateness monotonically increases from its
initial value (unity) to its asymptotic value. More string density
increases the anisotropy; more matter reduces anisotropy, but neither
can change an oblate ellipsoid into a prolate one.

The asymptotics for large $t$ are as follows:
\begin{eqnarray}
&&\e\sim \e_\i\,e^{-2(\l/3)^{\frac{1}{2}}(t-t_\i+\tau)},
\label{S:e-asympt}\\
&&a\sim(\l/3)^{\frac{1}{2}}(t-t_\i+\tau), \label{S:a-asympt}\\ &&b\sim
(\l/3)^{\frac{1}{2}}(t-t_\i+\tau)-\fr{1}{2}
\ln{\left[1+(\rho_\i+\e_\i)/\l\right]}+\ln{[1+F_\textrm{s}(0)]},\label{S:b-asympt}\\
&&\rho\sim\rho_\i[1+(\e_\i+\rho_\i)/\l]^{\frac{1}{2}}[1+F_\textrm{s}(0)]^{-1}
e^{-(3\l)^{\frac{1}{2}}(t-t_\i+\tau)},\label{S:rho-asympt}
\end{eqnarray}
where
\begin{equation}
\tau=\fr{1}{2}\int_{0}^{\e_\i}
\d\e\left\{\left(\fr{1}{3}\l\e^2\right)^{-\frac{1}{2}}
-\left[\fr{1}{3}\l\e^2+\fr{1}{3}\e_\i^{-\frac{3}{2}}(\rho_\i-2\e_\i)
\e^{\frac{7}{2}}+\e^3\right]^{-\frac{1}{2}}\right\}.\label{S:tau}
\end{equation}

\section{Walls}\label{S:walls}

Energy-momentum conservation gives $b=\ln{(\e_\i/\e)}$. Eq.~(\ref{A3})
is easily solved to give
\begin{equation}
a=\fr{2}{3}\ln \frac{e^{(3\l)^\frac{1}{2}(t-t_\i)}-\s}{1-\s}
-(\l/3)^\frac{1}{2}(t-t_\i),\label{W:a}
\end{equation}
where
\begin{equation}
\s=\frac{\left[1+(\r_\i+\e_\i)/\l\right]^{\frac{1}{2}}-1}
{\left[1+(\r_\i+\e_\i)/\l\right]^{\frac{1}{2}}+1}.\label{W:sigma}
\end{equation}
Using Eqs.~(\ref{A1}) and (\ref{W:a}) together with $\rho=\rho_\i
e^{-(2a+b)}$, we find
\begin{equation}
\e=\e_\i\g\frac{\left(\g^3-\s\right)^\frac{1}{3}}{\g^3+\s}F_\textrm{w}(\g)
,\label{W:e}
\end{equation}
where 
\begin{eqnarray} 
F_\textrm{w}(\g)=\left[\frac{\left(1-\s\right)^\frac{1}{3}}{1+\s}
+\frac{\r_\i}{2\l}\frac{\left(1-\s\right)^\frac{4}{3}}{1+\s}
\frac{\g^3-1}{\g^3+1}+\frac{3\e_\i}{2\l}\int_1^\g \d
x\,\frac{(x^3-\s)^{\frac{4}{3}}}{(x^3+\s)^2} \right]^{-1}
\end{eqnarray} 
and $\g=e^{(\l/3)^\frac{1}{2}(t-t_\i)}$.

Inspecting the above solution shows that space is prolate, $e^{a-b}\le
1$, and its prolateness monotonically increases from its initial value
(unity) to its asymptotic value. More wall density increases the
anisotropy; more matter reduces anisotropy, but neither can change an
prolate ellipsoid into a oblate one.

The asymptotics for large $t$ are as follows:
\begin{eqnarray}
&&\e\sim \e_\i\,e^{-(\l/3)^\frac{1}{2}
(t-t_\i)}F_\textrm{w}(\infty),\label{W:e-asympt}\\ &&a\sim
(\l/3)^\frac{1}{2}(t-t_\i)-\fr{2}{3}\ln{(1-\s)},\label{W:a-asympt}\\
&&b\sim (\l/3)^\frac{1}{2}(t-t_\i)-\ln
F_\textrm{w}(\infty),\label{W:b-asympt}\\
&&\rho\sim\rho_\i(1-\s)^{\frac{4}{3}}\,e^{-\mu(t-t_\i)}F_\textrm{w}(\infty).
\end{eqnarray}

\begin{figure}
\begin{center}
\includegraphics[width=12cm]{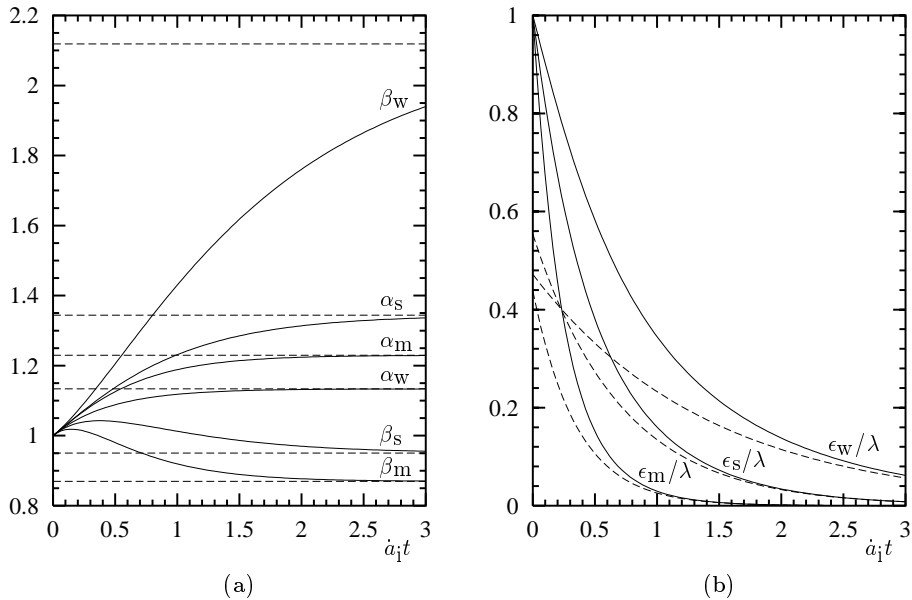}
\caption{(a) The ratios of the asymmetric scale factors to the
symmetric scale factors, $\alpha=e^{a-c}$ and $\beta=e^{b-c}$, where
$c=(\fr{1}{3}\l)^{\frac{1}{2}}(t-t_\i)$. In all three cases the same
initial conditions are used. Solid lines are exact solutions and
dashed lines are asymptotics for large $t$. (Matter is not included.)
(b) The ratio $\e/\l$ for the same three cases and initial conditions
as in (a).}\label{fig-1}
\end{center}
\end{figure}

\section{Conclusions}

To apply the results of this paper, it will be necessary to consider
how the spectrum of density perturbations are effected by asymmetric
expansion.  Since perturbations get laid down by quantum fluctuations
and then asymmetrically expanded in our model, any initial spherical
perturbation becomes ellipsoidal. After a while the expansion becomes
spherically symmetric, but as long as perturbations remain outside the
horizon they stay ellipsoidal.  Only after they enter our horizon will
they be able to adjust (they will probabily start to oscillate between
prolate and oblate with frequency that depends on size and
overdensity). So if the perturbations are just entering at last
scattering they should be ellipsoidal. The smaller they are at last
scattering, the more they have oscillated and if damped, the closer to
spherical they should be. Hence, the large perturbations
(corresponding to small $l$) will have a better memory of our
eccentric phase. This would give what seems to be hinted at in the
WMAP observations--more distortion of the low $l$ modes.

To summarize, what we need are modes that expanded eccentrically to be
entering the horizon at the time of last scattering and then to feed
into the Sachs-Wolfe calculation. This is a most interesting and
challenging calculation, since it requires a full reanalysis of the
density perturbations in eccentric geometry~\cite{perturbations}.

\section*{Acknowledgments}

This work is supported in part by U.S. DoE grant number
DE-FG05-85ER40226.


\begin{thebibliography}{99}



\bibitem{Berera:2003tf} A.~Berera, R.~V.~Buniy and T.~W.~Kephart,
arXiv:hep-ph/0311233.

\bibitem{long-version} A.~Berera, R.~V.~Buniy and T.~W.~Kephart,
``Asymmetric inflation: exact solutions,'' in preparation.

\bibitem{Bennett:2003bz} C.~L.~Bennett {\it et al.},
Astrophys.\ J.\ Suppl.\ {\bf 148}, 1 (2003).


\bibitem{Spergel:2003cb} D.~N.~Spergel {\it et al.},
Astrophys.\ J.\ Suppl.\  {\bf 148}, 175 (2003).

\bibitem{Hinshaw:2003ex} G.~Hinshaw {\it et al.},
Astrophys.\ J.\ Suppl.\ {\bf 148}, 135 (2003).

\bibitem{Tegmark:2003ve} M.~Tegmark, A.~de Oliveira-Costa and
A.~Hamilton,
arXiv:astro-ph/0302496.

\bibitem{deOliveira-Costa:2003pu} A.~de Oliveira-Costa, M.~Tegmark,
M.~Zaldarriaga and A.~Hamilton,
arXiv:astro-ph/0307282.

\bibitem{Smoot:1992td} G.~F.~Smoot {\it et al.},
Astrophys.\ J.\  {\bf 396}, L1 (1992).

\bibitem{Kogut:1996us} A.~Kogut {\it et. al},
arXiv:astro-ph/9601060.

\bibitem{Kronberg:1993vk} P.~P.~Kronberg,
Rept.\ Prog.\ Phys.\  {\bf 57}, 325 (1994).

\bibitem{Hindmarsh:1994re} M.~B.~Hindmarsh and T.~W.~Kibble,
Rept.\ Prog.\ Phys.\ {\bf 58}, 477 (1995).

\bibitem{Wick:2000yc} S.~D.~Wick, T.~W.~Kephart, T.~J.~Weiler and
P.~L.~Biermann,
Astropart.\ Phys.\ {\bf 18}, 663 (2003).

\bibitem{Taub:1950ez} A.~H.~Taub,
Annals Math.\ {\bf 53}, 472 (1951).

\bibitem{Savvidy:1977as} G.~K.~Savvidy,
Phys.\ Lett.\ B {\bf 71}, 133 (1977).

\bibitem{Matinian:1976mp} S.~G.~Matinian and G.~K.~Savvidy,
Nucl.\ Phys.\ B {\bf 134}, 539 (1978).

\bibitem{Vachaspati:nm} T.~Vachaspati,
Phys.\ Lett.\ B {\bf 265} (1991) 258.

\bibitem{Enqvist:1994rm} K.~Enqvist and P.~Olesen,
Phys.\ Lett.\ B {\bf 329}, 195 (1994).

\bibitem{Berera:1998hv} A.~Berera, T.~W.~Kephart and S.~D.~Wick,
Phys.\ Rev.\ D {\bf 59}, 043510 (1999).

\bibitem{perturbations} A.~Berera, R.~V.~Buniy and T.~W.~Kephart,
``Asymmetric inflation: density perturbations,'' in preparation.




\end{thebibliography}
\end{document}